\documentclass[superscriptaddress, showpacs,showkeys,twocolumn,showpacs,amsmath,amssymb]{revtex4}
\usepackage{amsmath}
\usepackage{graphicx}
\usepackage{dcolumn}
\usepackage{amsfonts}
\usepackage{amssymb}
\usepackage[colorlinks=true,citecolor=cyan, linkcolor=cyan, filecolor=cyan,urlcolor=cyan]{hyperref}
\usepackage{mymacros}
\usepackage{verbatim}

\begin{document}
\title{Principal Flow Patterns across renewable electricity networks}

\author{Fabian Hofmann}
\affiliation{%
 Frankfurt Institute for Advanced Studies, Ruth-Moufang-Stra{\ss}e 1, 60438~Frankfurt~am~Main, Germany
}
\author{Mirko Sch\"{a}fer}
\affiliation{%
Department of Sustainable Systems Engineering (INATECH), University of Freiburg, Emmy-Noether-Stra{ss}e 2, 79110~Freiburg, Germany
}
\affiliation{%
Department of Engineering, Aarhus University, Inge Lehmanns Gade 10, 8000~Aarhus~C, Denmark
}
\author{Tom Brown}
\affiliation{%
Institute for Automation and Applied Informatics, Karlsruhe Institute of Technology, 76344~Eggenstein-Leopoldshafen, Germany
}
\affiliation{%
 Frankfurt Institute for Advanced Studies, Ruth-Moufang-Stra{\ss}e 1, 60438~Frankfurt~am~Main, Germany
}
\author{Jonas H\"{o}rsch}
\affiliation{%
Institute for Automation and Applied Informatics, Karlsruhe Institute of Technology, 76344~Eggenstein-Leopoldshafen, Germany
}
\affiliation{%
 Frankfurt Institute for Advanced Studies, Ruth-Moufang-Stra{\ss}e 1, 60438~Frankfurt~am~Main, Germany
}
\author{Stefan Schramm}
\affiliation{%
 Frankfurt Institute for Advanced Studies, Ruth-Moufang-Stra{\ss}e 1, 60438~Frankfurt~am~Main, Germany
}
\author{Martin Greiner}
\affiliation{%
Department of Engineering, Aarhus University, Inge Lehmanns Gade 10, 8000~Aarhus~C, Denmark
}


\maketitle
%
%
\section{Introduction}
%
%
Dimensional reduction is a widely used method in multivariate statistical analysis. The spatio-temporal patterns in large complex systems with high dimensional data become more transparent once they are reduced to the lower dimensional space where their important interactions take place. The general idea of Principal Component Analysis (PCA) is to reduce a multivariate system of dimension $N$ to a set of $\mathcal{K}$ axes which represent as much variance of the original system as possible. Co-varying directions within the original data are exploited in order to build a smaller set of new, independent variables, the so-called Principal Components. This method is useful for high dimensional systems, such as those often found in meteorology that have a fine spatial and temporal resolution. The European power system represents another example, where the high number of nodes, transmission lines and time-steps represents a challenge for spatio-temporal analysis.

There have been several applications of PCA to power systems in the literature. In~\cite{burke_study_2011} a dimensional reduction of Irish wind power generation data to its Principal Components is investigated. An operational application is given in~\cite{deladreue_using_2002}, where line congestions in a small power system are located using PCA. Another operational application can be found in~\cite{xiao-fei_power_2016}, where PCA is combined with machine learning in order to predict load data. A more general approach is taken in~\cite{raunbak_principal_2017}, where the leading principal components of the residual load at nodes in a highly renewable European system are associated to typical recurring weather patterns. This study was realised using a coarse network representation with one node per country, which is known to be not sufficient to resolve the correlation length of wind systems (around 250-600~km \cite{npg-15-803-2008,Widen2011,st._martin_variability_2015}).

In the present article we apply PCA to a spatially detailed model of a highly renewable future European electricity system. The influence of the spatial resolution of the model for the dynamical injection and flow patterns is investigated by applying a clustering algorithm to the detailed original data. The PCA is applied to the injection patterns representing the nodal imports and exports, and to the power flow patterns taking place on the network model representing the transmission grid. Arguments from matrix theory and complex networks physics are used to explain the surprisingly low number of principal flow patterns found in this system.

The article is structured as follows. After the introduction, the network model of a highly renewable European electricity system, the power flow equations, and some fundamentals on principal component analysis are presented. Subsequently we first apply PCA to the injection and then to the power flow patterns occurring in the system. The following theoretical section 
explains the connection between injection patterns, network topology, and power flow patterns. A conclusion and outlook is given in the last section.
%
%
\section{Modeling}
%
%
The European electricity system model used in this study was initially presented by H\"orsch and Brown in~\cite{Hoersch2017}, where transmission and generation optimisation for different spatial scales were studied. Using a k-means clustering methodology, the network with 4653 nodes is gradually reduced to the sizes $N \in$ \{1024, 724, 502, 362, 256, 181, 128, 90, 64, 45, 37\}. The model is implemented in the free software `Python for Power System Analysis (PyPSA)' Version 0.12.0 \cite{PyPSA}, which was developed at the Frankfurt Institute for Advanced Studies (FIAS). The set of buses and lines with corresponding capacities is derived from the online ENTSO-E Interactive
Transmission Map \cite{interactive} using the GridKit extraction
\cite{wiegmans_2016_55853}. In the ENTSO-E area, all transmission lines with voltage at or above 220~kV and all HVDC lines are included within the model. For simplicity's sake, we convert all lines (AC and DC) to 380~kV AC lines.

Using the Aarhus Renewable Energy Atlas~\cite{andresen2015}, weather-dependent solar and wind power potential time-series are derived from historical weather data (time span 2011-2014, with hourly resolution and a spatial resolution of $40\times 40~\textrm{km}^{2}$). The hourly electricity demand profiles for each country for 2011-2014 are
taken from the European Network of Transmission System Operators for
Electricity (ENTSO-E) website~\cite{entsoe_load}. The spatial
distribution of load within each country is obtained using a regionalisation procedure based on the GDP and the population statistics for the NUTS3 regions, which are an EU geocode standard for referencing the subdivision of countries for statistical purposes.

The renewable power generation $ g_n^R(t) $ at each node $n$ is composed of wind and solar power generation, that is $g_n^R(t) = g_n^W (t) + g_n^S(t)$. For each country $c$ we define $ \alpha_c $ as the proportion of the load that is on average covered from wind power generation. Assuming an average renewable penetration of 100\%, the remaining proportion $ (1-\alpha_c) $ is the solar share of the country $c$, that is 
\begin{align}
	\alpha_c = \dfrac{\sum_{n \in \text{c}} \< g_n^W \>}{\sum_{n \in \text{c}} \<l_n\>}, \;\;\; (1-\alpha_c) = \dfrac{\sum_{n \in \text{c}} \< g_n^S \>}{\sum_{n \in \text{c}} \<l_n\>}~,
\end{align}
where $ l_n (t)$ is the nodal electricity demand (load) and $\< \cdot \>$ the average over time~$t$. The wind and solar power generators are distributed proportionally to the product of capacity factor and installable capacity at the node $ n $, with protected sites and offshore regions with more than $50\,$m water depth not taken into account.    
The set of wind and solar shares $\{\alpha_{c}\}$ is chosen such that the total sum of the nodal renewable generation variances is minimised, which reduces weather-driven fluctuations over the whole system. This minimisation approach leads to a heterogeneous composition of renewable generation capacities, with more wind power generation in North Europe and more solar power generation in southern countries.

The nodal mismatch $ \Delta_n (t) $ is defined as the difference of renewable generation and load, $g_{n}^{R}(t)-l_{n}(t)$. In order to assure nodal balance, the mismatch has to be compensated by power generation management or power transfers,
\begin{align}
\Delta_n (t) = g^R_n(t) - l_n(t) = b_n(t) + p_n(t)~.
\label{eq:nodal_balancing}
\end{align}
The right hand side consists of the nodal power generation balancing $b_n$, combining curtailment and backup power at node $ n $, and the net power injection $p_n$ into the network. Note that these terms can be expanded by other technologies, for instance storage or coupling to other sectors~\cite{brown_synergies_2018,SCHLACHTBERGER2017469}. We apply a balancing scheme $b_n$ which sets the curtailment of the renewable generation proportional to the actual nodal energy excess, and the backup energy proportional to the mean load~\cite{rodriguez_localized_2015}:
\begin{align}
b_n(t)  = \dfrac{\Delta_{n}^+(t)}{\sum_m \Delta_{m}^+(t)} \Delta^+(t) - \dfrac{\<l_n\>}{\sum_m \<l_m\>} \Delta^-(t)
\label{eq:nodal_balancing}
\end{align}
Here we use the definitions $\Delta_{n}^{\pm}(t) = \text{max} \left(0, \pm\Delta_{n} (t)\right)$ and $\Delta^{\pm}(t)=\sum_{n}\Delta_{n}^{\pm}(t)$.
The power flow  is calculated using the unconstrained linear power flow approximation \cite{van_hertem_usefulness_2006}, which maps the injection pattern vector $\mathbf{p}(t)$  linearly to the flow vector $\mathbf{f}(t)$,
\begin{align}
\mathbf{f}(t) &= \textbf{H}\; \mathbf{p}(t)~.
\label{eq:flow_equation}
\end{align}
Here $\mathbf{H} = \mathbf{\Omega}\, \mathbf{K}^T\, \mathbf{B}^{+}$ is the power transmission distribution factor (PTDF) matrix of size $L \times N$, with $\mathbf{\Omega}: L\times L$ containing the inverse of the line reactants $x_l^{-1}$ on its diagonal, $\mathbf{K}:N \times L$ being the incidence matrix, and $\mathbf{B}^{+}$ the Moore-Penrose pseudo inverse of the nodal susceptance matrix.
%
\section{Power Injection PCA}
\label{sec:injection_PCA}
%
%
For a time-dependent variable $\mathbf{x}(t)$, the Principal Axes $\{ \PA_k \}$ are given by the eigenvectors of the covariance matrix $\mathbf{\Sigma}^{\text{x}} = \mathbf{\Sigma}(\mathbf{x})=\text{cov}(\mathbf{x}, \mathbf{x})$. Through normalisation of the covariance matrix $\mathbf{\Sigma}/\text{tr}(\mathbf{\Sigma})$ the sum of the eigenvalues $\lambda_k$ is set to one,
\begin{align}
\tilde{\lambda}_{k}=\frac{\lambda_{k}}{\sum_{k'}\lambda_{k'}}=\frac{\lambda_{k}}{\text{tr}(\mathbf{\Sigma})}~.
\end{align}
The set $ \{ \PA_k \} $ defines a new basis where $ \lambda_k $ indicates the proportion of variance occurring on the axis $\PA_k$ \cite{jolliffe_principal_2002}.We assume all eigenvectors to be ordered according to the size of their eigenvalue, $\lambda_1$ being the largest eigenvalue.
The amplitude $\beta_k(t)$ is defined as the projection of the mean-free data onto the axis, \textit{i.e.} $ \beta_k(t) = (\mathbf{x}(t)-\<\mathbf{x}\>)\, \PA_k $. The Principal Components (PC) are given by the product $\tilde{\lambda}_k\, \PA_k$. We set $ \mathcal{K} $ as the number of axes needed to explain 95\% of the total variance, that is $\sum_{k=1}^\mathcal{K} \tilde{\lambda}_k \geq 0.95$. For convenience, in the following we refer to both the Principal Components and the Principal Axes as the Principal Components (PCs) or patterns of the system.

\begin{figure}
	\includegraphics[width=\linewidth]{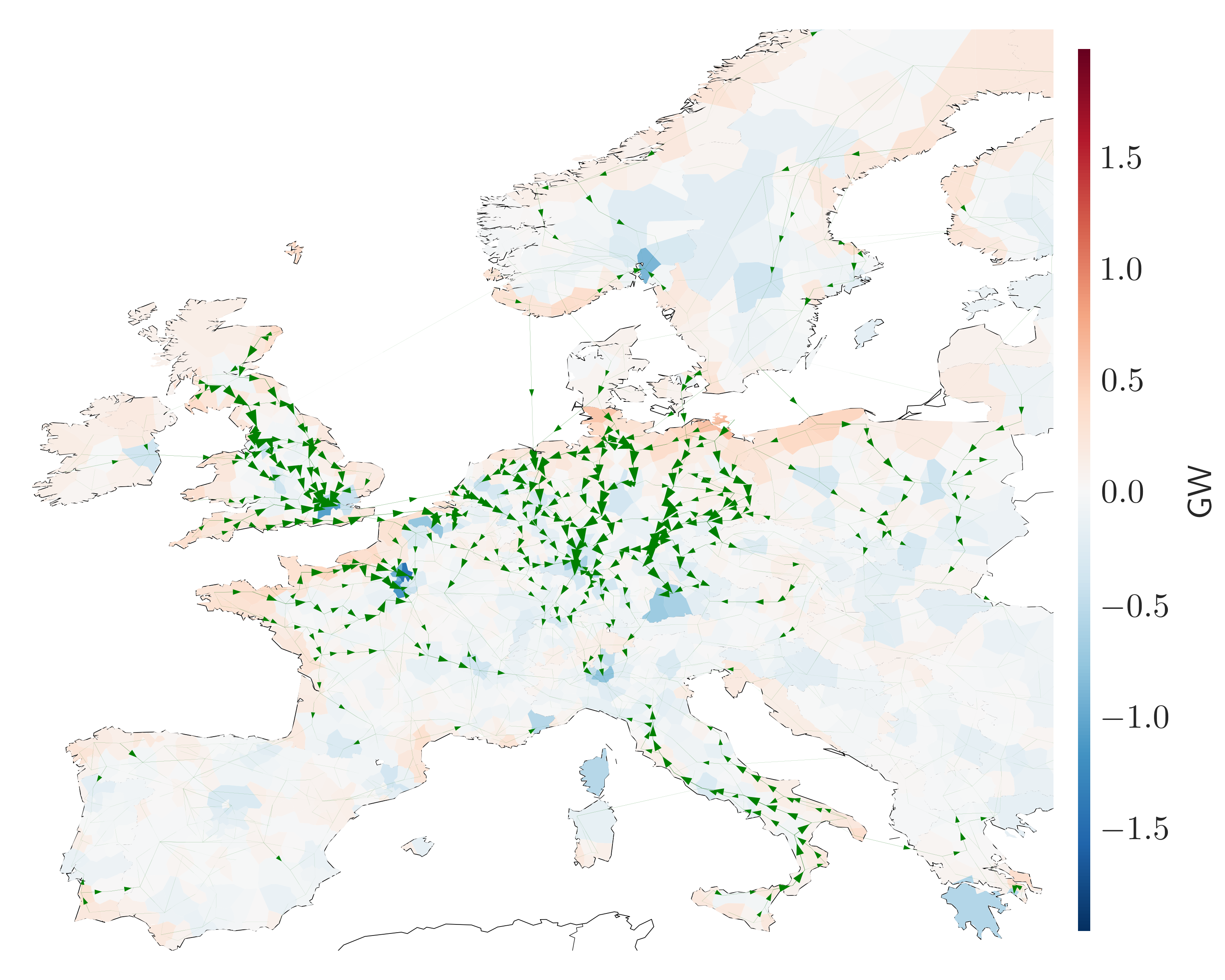}
	\caption{Average injection $\langle\mathbf{p} \rangle $ and average power flow $\langle \mathbf{f}\rangle$. In contrast to densely populated areas which show significant power imports, coastal areas serve as feed-in sources. For clearer visualisation, only the upper $30\%$ quantile of the power flows is displayed. Unless stated otherwise, all figures in the article show results for network size  $N\,=\,1024$.}
	\label{fig:injection_mean_pre6-1024-harm-alpha_inhom-064}
\end{figure}

\begin{figure*}
	\includegraphics[width=\linewidth]{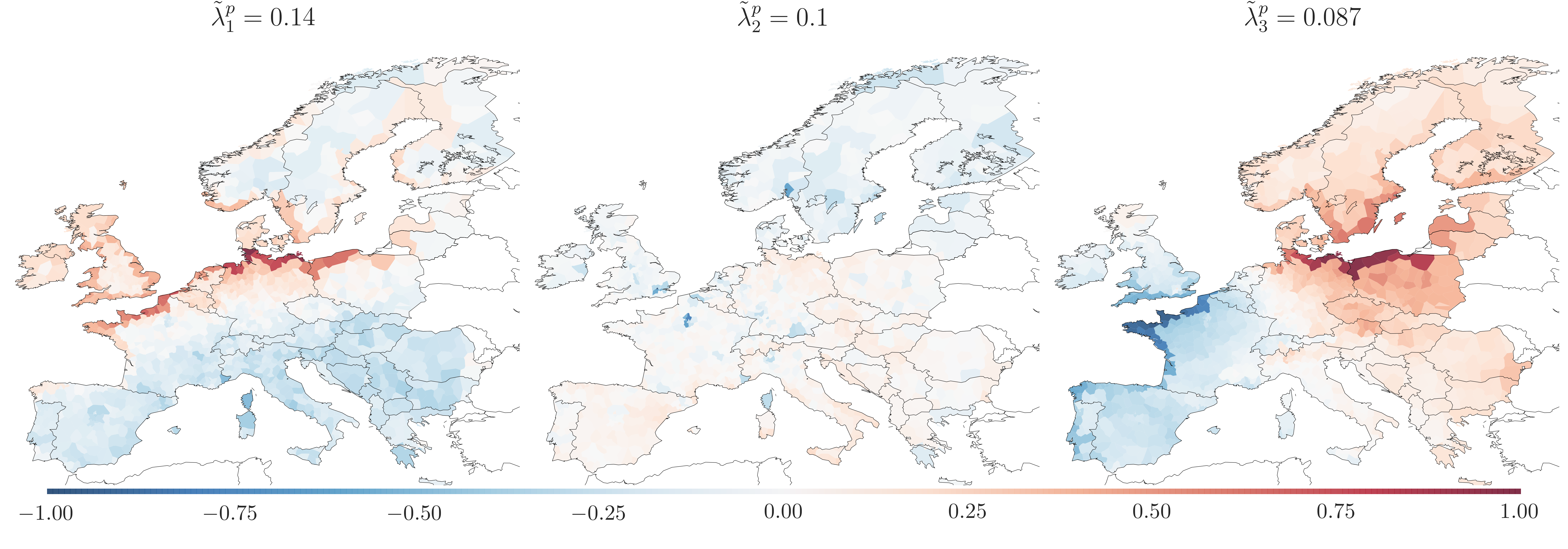}
	\caption{First three injection principal components $\PAp_k$ and normalised eigenvalues $\tilde{\lambda}^{p}_{k}$. These principal injection patterns reveal a bipole behavior north-south and east-west with an emphasis on the power generation centers along the North Sea coast.}
	\label{fig:injection_components_pre6-1024-harm-alpha_inhom-064}
\end{figure*}

\begin{figure}
	\includegraphics[width=\linewidth]{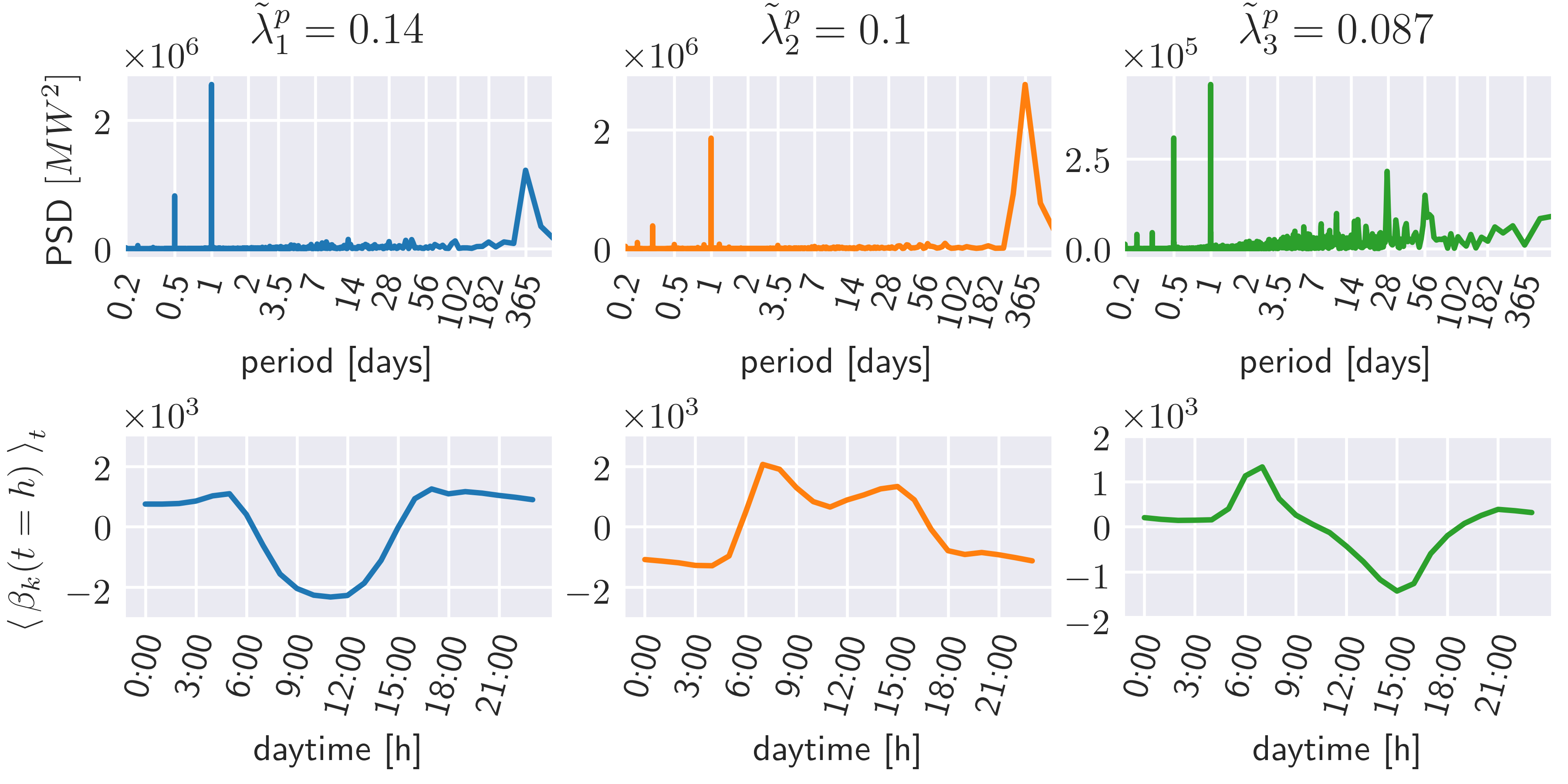}
	\caption{Power Spectra Densities (PSD) and daytime profiles of the first three injection PCs. The PCs have strong diurnal cycles which are shown in the daytime profiles, but additionally reveal seasonal cycles.}
	\label{fig:injection_fourrier_daytime_pre6-1024-harm-alpha_inhom-064}
\end{figure}

Figure~\ref{fig:injection_mean_pre6-1024-harm-alpha_inhom-064} shows the mean injection $\langle\mathbf{p}\rangle$ and the mean flow $\langle\mathbf{f}\rangle$ for the model of the renewable European electricity system. Cities like Paris and London with strong power imports are highlighted in contrast to coastal areas in north Europe which accumulate the feed-ins of the offshore regions. In Germany and the United Kingdom, a power injection gradient from north to south is apparent. Figure~\ref{fig:injection_components_pre6-1024-harm-alpha_inhom-064} shows the spatial distribution of the first three PCs, with fig.~\ref{fig:injection_fourrier_daytime_pre6-1024-harm-alpha_inhom-064} illustrating their time-dependent behavior. The latter shows the Power Spectra Densities (PSD), which is the Fourier-transform of the amplitude $\beta_k$, and the daytime profiles, giving the average value of $\beta_k$ at daytime $h$. 
%
%
%
%
Similar to the mean, the first component~$ \PA_1 $ shows a strong (co-)variation of the coastal areas along the North Sea with an anti-aligned behavior of South Europe. This component is strongly negative during daytime, as power is fed-in in south Europe due to solar power excess. The North supplies wind power during evening and night. Similar to the findings of~\cite{raunbak_principal_2017} this PC indicates a bipole between north and south, which occurs on a daily basis.
For the second component, most of middle and South Europe have slightly positive injection, the North and especially Paris, London and Oslo reveal negative power injection. The third PC reflects a bipole between east and west Europe, which occurs in morning and evening time. This injection behaviour can be associated with the effect of sunrise and sunset, as solar power is generated during morning in east where the sun is already risen up, or during evening in west where it has not set down yet. This time shifted solar power production is aligned with strong injections along the coast in the North.
%
%
The PSD in fig.~\ref{fig:injection_fourrier_daytime_pre6-1024-harm-alpha_inhom-064} reflects the strong relation of the weather-driven energy supply to the climatic rhythms: The most prominent periodic sequence of the first three PCs is the diurnal cycle followed by half-day and seasonal cycles. 
%
\begin{figure}
	\includegraphics[width=\linewidth]{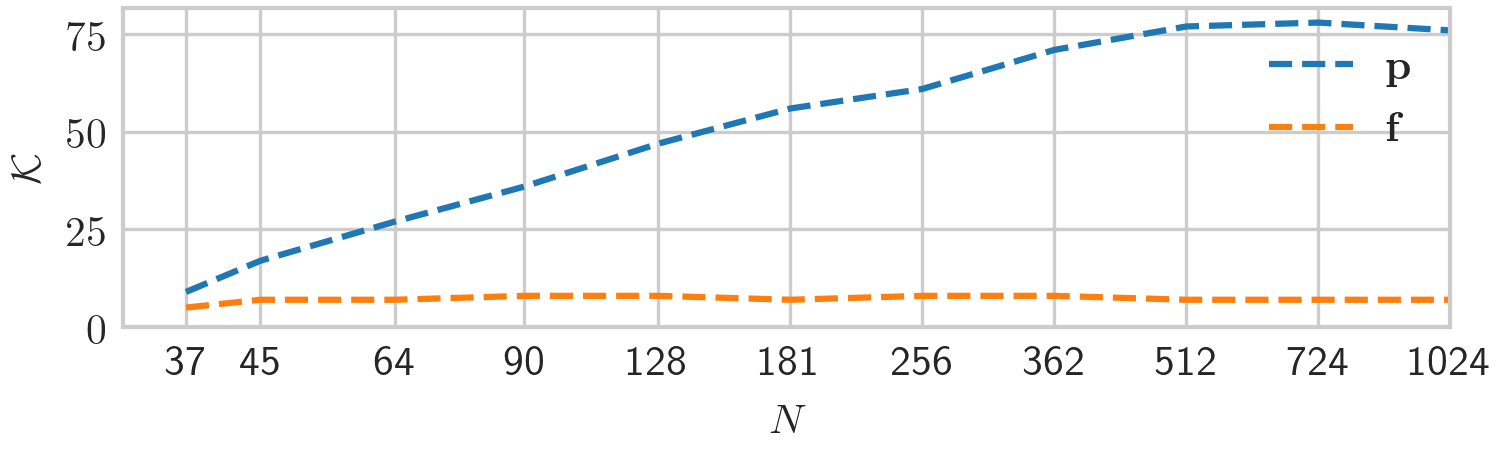}
	\caption{Number of relevant components $ \mathcal{K} $ as a function of the network size $N$. Whereas the number of relevant injection patterns increases with increasing number of nodes, the total flow can be accurately decomposed to a small set of $ \mathcal{K}$\,$\approx$\,8 principal components even for large networks.}
	\label{fig:K_size_dependence}
\end{figure}
%

Figure~\ref{fig:K_size_dependence} shows the number of relevant injection patterns $ \mathcal{K} $ with $\sum_{k}^{\mathcal{K}}\tilde{\lambda}_{k}^{p}\geq 0.95$ as a function of the network size $N$. With increasing spatial resolution, $ \mathcal{K} $ at first almost scales linearly with log($N$). However, at $N$\,=\,512 the number of relevant injection patterns saturates with $ \mathcal{K}$\,$\approx$\,76. With $ N $\,=\,512 the average next-neighbour-distance is $ \xi = (A/N)^{-1} \approxeq 138$\,km, where $A$ indicates the total network area of $9.834\cdot 10^6$ \,km$^2$ (including offshore regions). When further increasing $ N $, that is decreasing $\xi$, additional nodes rather interpolate between their next neighbours, as co-varying regions (cohesive regions of one color in~fig. \ref{fig:injection_components_pre6-1024-harm-alpha_inhom-064}) of the most relevant PCs do not exceed an area of $\xi^2$. According to the typical scale of weather patterns, one would expect $ \mathcal{K} $ to saturate earlier (for $\xi = \xi_{W} = 273$\,km, where $ \xi_{W} $ denotes the wind correlation length for the northern hemisphere stated in \cite{st._martin_variability_2015}). However, localised effects of load and backup generation bring the relevant spatial scale down to $ \xi \approxeq 138$\,km.
%
%
\section{Power Flow PCA}
%
%
PCA can also be applied to the ensemble of power flow patterns $\mathbf{f}(t)$ resulting from the injection patterns $\mathbf{p}(t)$ studied in the previous section. Given that these flow patterns $\mathbf{f}(t) = \textbf{H}\; \mathbf{p}(t)$ follow from the injection patterns linearly through matrix multiplication, one would naively expect a similiar number of PCs. Surprisingly, fig.~\ref{fig:K_size_dependence} shows that almost independently from the network size $N$ about $ \mathcal{K} =8$ components are sufficient to approximate the dynamics of the total network flow.

\begin{figure*}
	\includegraphics[width=\linewidth]{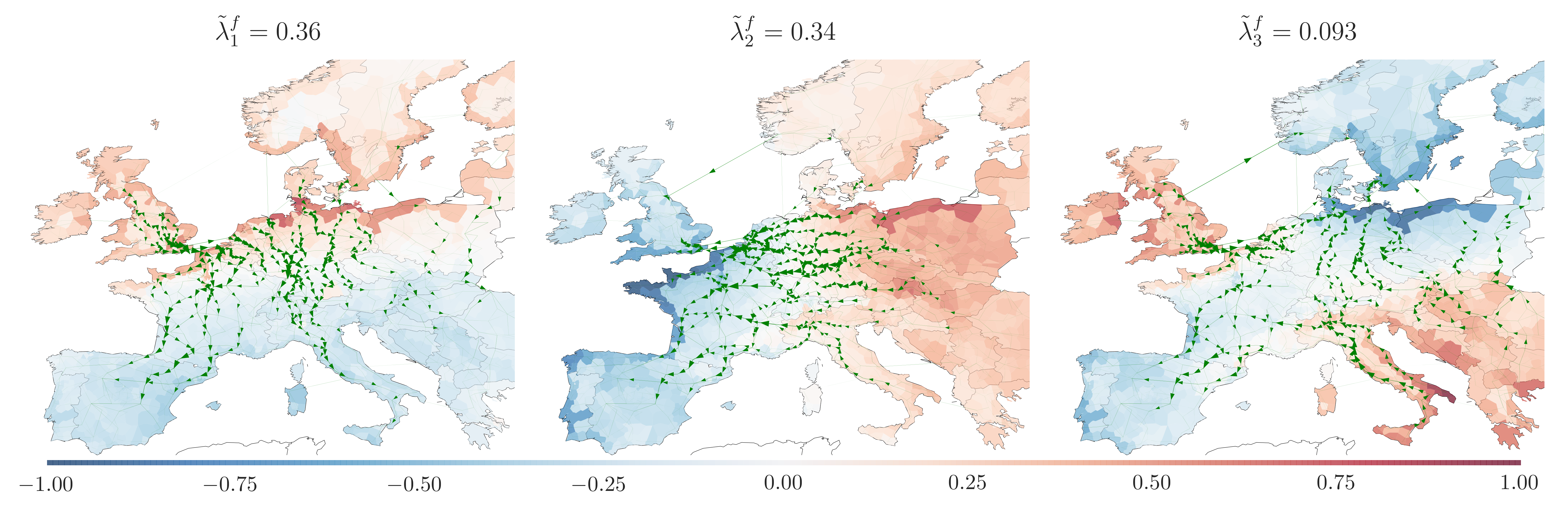}
	\caption{First three principal flow patterns of a highly renewable European electricity network. For clearer visualisation only the upper $20\%$-quantile of the flows is displayed. The first two principal components $\PAf_1$ and $\PAf_2$ show an overall orientation along the North-South and East-West axis, respectively. These long distance flow PCs cover already 70\% of the flows total variance. The third component shows flows from the North-West and South-East to the South-West and North-East. The  colouring of the map represents the normalised eigenvectors $\mathbf{v}_{k}$ of the matrix $\mathbf{M}$ corresponding to these principal flow patterns, see eqs.~(\ref{eq:ev_M}) and~(\ref{eq:eigenvector_connection}).}
	\label{fig:flow_components_pre6-128-harm-alpha_inhom-071}
\end{figure*}

Through eq.~(\ref{eq:flow_equation}) the mean flow is mapped to the mean injection, $\langle\mathbf{f}\rangle = \mathbf{H} \langle\mathbf{p}\rangle $, which along with the average injection pattern $\langle\mathbf{p}\rangle$ is displayed in fig.~\ref{fig:injection_mean_pre6-1024-harm-alpha_inhom-064}. One observes that the average power flows are directed from the main power sources along the North Sea coast to the load centers with dense population. In fig.~\ref{fig:flow_components_pre6-128-harm-alpha_inhom-071} we display the first three principal flow patterns which already are associated with 79\% of the flow's total variance. The two first flow principal components~$\PAf_1$ and~$\PAf_2$ correspond to long-distance flows mainly oriented along the North-South and East-West axis, respectively. The third PC $\PAf_3$ displays main currents from the North-West and South-East to the South-West and North-East.
%
%
%
\begin{figure}
	\includegraphics[width=\linewidth]{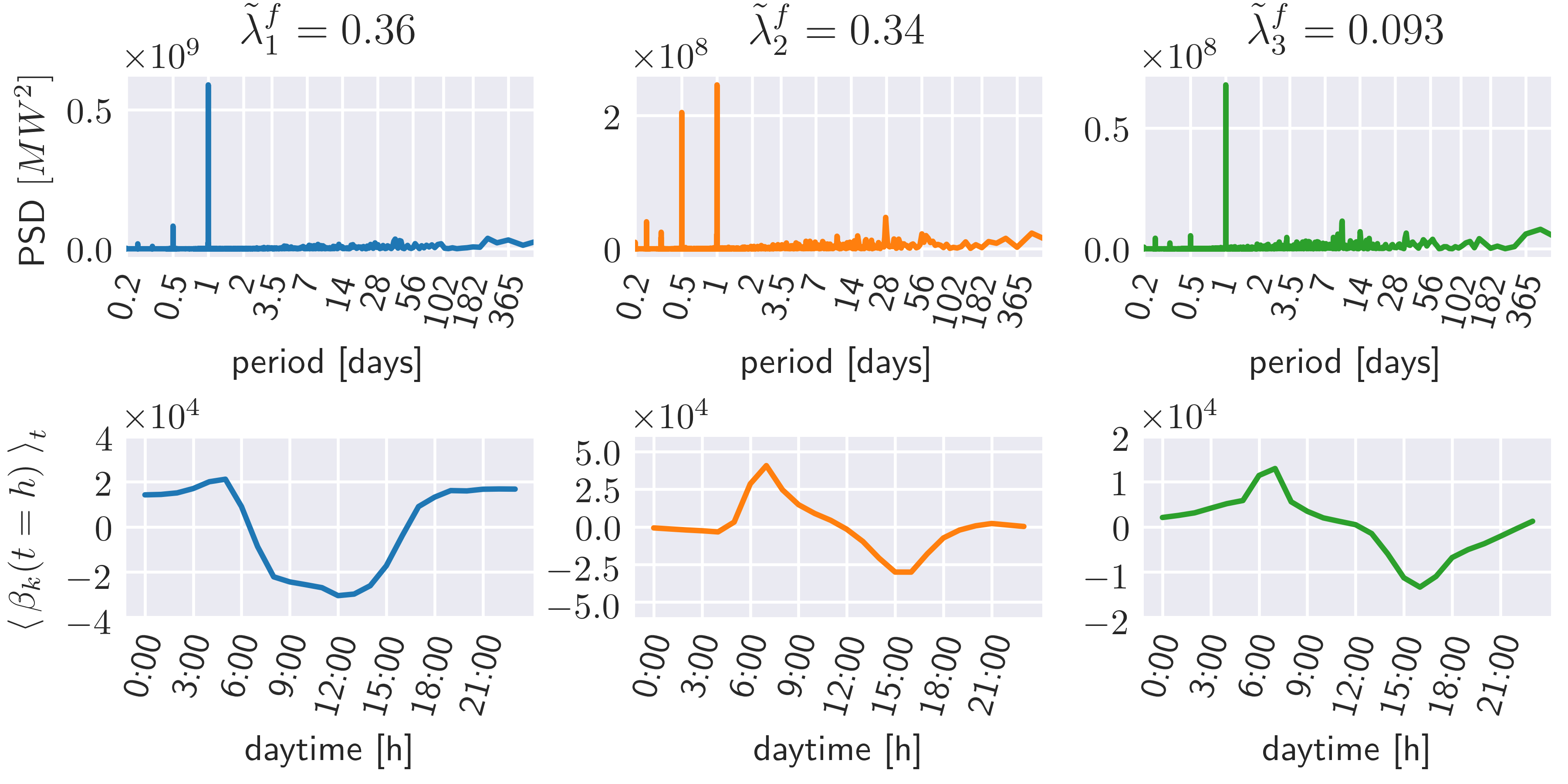}
	\caption{Power Spectra Densities (PSD) and daytime profiles of the first three power flow PCs. As for the injection patterns, one observes predominantly diurnal, but also seasonal cycles.}
	\label{fig:flow_fourrier_daytime_pre6-128-harm-alpha_inhom-064}
\end{figure}
%
The Power Spectra Densities of the principal flow components in fig.~\ref{fig:flow_fourrier_daytime_pre6-128-harm-alpha_inhom-064} show some similarity to the ones observed for the principal injection patterns in fig.~\ref{fig:injection_fourrier_daytime_pre6-1024-harm-alpha_inhom-064}, with the main periodicity given by the diurnal and the half-day cycle.
%
%
\section{Theoretical Relation of Power Injection PCA to Flow PCA}
%
%
In this section, we provide a theoretical explanation for the strong concentration of principal flow components shown in fig.~\ref{fig:K_size_dependence}. As a first step, it is crucial to recognise that the flow data $\mathbf{f}(t)$ contains only $N-1$ relevant PCs with a non-zero eigenvalue. Applying the covariance operator to eq.~(\ref{eq:flow_equation}) leads to
\begin{align}
\covf &= \mathbf{\Sigma}(\mathbf{H} \; \mathbf{p}(t)) = \mathbf{H} \; \covp\; \mathbf{H}^T~,
\label{eq:covariance_flow}
\end{align}
from which after normalisation the principal flow components $\PAf_k$ are extracted via matrix diagonalisation. The flow covariance matrix $\covf$ is of size $L \times L$, however through the mapping~(\ref{eq:covariance_flow}) its rank is
\begin{align}
\rk{\covf}=\text{min}\left(\rk{\mathbf{H}}, \rk{\covp} \right) = N-1~.
\end{align}
The nullspace of $\covf$ with dimension $\text{nul}(\covf)$ $=L-N+1$ is spanned by eigenvectors with zero eigenvalue corresponding to non-physical flow patterns in the network, which result from the $C$ cycles in the system~\cite{HORSCH2018126}. 

The remaining relevant $N-1$ PCs with non-zero eigenvalue span the image of the flow covariance matrix $\text{im}(\covf)$. In the following we will show that these $L$-dimensional eigenvectors are related to $N$-dimensional eigenvectors of a $N\times N$ matrix determined by the injection pattern and the network structure. For this purpose we define the $N \times N$ `topological matrix' $\mathbf{T} = \mathbf{H}^T\,\mathbf{H}$, which as a product of the PTDF matrix $\mathbf{H}$ with itself only depends on the network representation of the power grid. Multiplication with the injection covariance matrix $\covp$ yields the $ N \times N $ matrix $\mathbf{M} =  \covp \, \mathbf{T}$.
Both matrices $\mathbf{T}$ and $\mathbf{M}$ have rank $N-1$. It can be shown that the eigenvectors of $\mathbf{M}$ determine the principal flow components:  Let $\mathbf{v}_k$ be the normalised eigenvectors and $ \eta_k $ the corresponding eigenvalues of $ \mathbf{M} $. By multiplying both sides of eq.~(\ref{eq:covariance_flow}) with $ \mathbf{H} \, \mathbf{v}_k $, we obtain
\begin{align}
\covf \, \mathbf{H} \, \mathbf{v}_k  &= \mathbf{H} \, \covp\, \mathbf{H}^T \, \mathbf{H} \, \mathbf{v}_k 
	= \mathbf{H} \, \mathbf{M}\, \mathbf{v}_k = \eta_k \, \mathbf{H} \, \mathbf{v}_k~.
\label{eq:ev_M}
\end{align}
Thus, the flow PCs with non-vanishing eigenvalue $\{\PAf_k | \lambda_k^f > 0\}$ are given by $ \{\mathbf{H} \, \mathbf{v}_k | \eta_k >0 \}$. The image of $ \covf $ and the complete set of relevant principal flow components is given by the linear mapping of eigenvectors of $ \mathbf{M}$,
\begin{align}
	\PAf_k = \dfrac{\mathbf{H}\,\mathbf{v}_k}{\norm{\mathbf{H}\, \mathbf{v}_k}}
\quad , \quad
\tilde{\lambda}^f_k = \frac{\eta_k}{\sum_{k}\eta_{k}} =  \frac{\eta_k}{\text{tr}(\mathbf{M})}~.
\label{eq:eigenvector_connection}
\end{align}
This reduces the calculation of the principal flow components to the eigen-decomposition of $ \mathbf{M} $. Figure~\ref{fig:flow_components_pre6-128-harm-alpha_inhom-071} shows the first three eigenvectors $\mathbf{v}_{k}$ of $\mathbf{M}$ along with the resulting flow PCs $\PAf_k \propto \mathbf{H}\,\mathbf{v}_{k}$. One observes a similarity between $\mathbf{v}_{1}$ and the first injection PC $\PA_1$, and between $\mathbf{v}_{2}$ and the third injection PC $\PA_3$. This finding explains the resemblance of the time evolution of the corresponding amplitudes as displayed in fig.~\ref{fig:injection_fourrier_daytime_pre6-1024-harm-alpha_inhom-064} and~\ref{fig:flow_fourrier_daytime_pre6-128-harm-alpha_inhom-064}.

 Despite $\mathbf{M}$ consists of a simple multiplication of two symmetric matrices, $ \mathbf{\covp} $ and $ \mathbf{T} $, in general it is non-trivial to connect its eigenvalues to the eigenvalues of its factors. Nevertheless, using majorisation it is possible to define lower and upper bounds for partial sums over the ordered sequence of $ \tilde{\lambda}^f_k $~\cite{bhatia_matrix_1997}.  We denote $ \Lambda^\downarrow(\mathbf{X})$ as the vector of non-zero eigenvalues of $ \mathbf{X} $ in decreasing order, and $ \boldsymbol{\Lambda}^\uparrow(\mathbf{X})$ likewise but in increasing order. From Prob.III.6.14 (corrected) in \cite{bhatia_matrix_1997} it follows
\begin{align}
  \sum_{k=1}^{K}\lambda_{k}^{f} & \geq \sum_{k=1}^{K}\left(\boldsymbol{\Lambda}^\downarrow(\mathbf{\covp}) \circ \boldsymbol{\Lambda}^\uparrow(\mathbf{T})\right)^{\downarrow}_{k}~,
  \label{eq:boundaries1}
\\
  \sum_{k=1}^{K}\lambda_{k}^{f} & \leq
  \sum_{k=1}^{K}\left(\boldsymbol{\Lambda}^\downarrow(\mathbf{\covp}) \circ \boldsymbol{\Lambda}^\downarrow(\mathbf{T})\right)^{\downarrow}_{k}~,
\label{eq:boundaries2}
  \end{align}
where $ (\circ) $ denotes the elementwise product. Here we have assumed that the eigenvalues $\lambda_{k}^{f}$ of $\mathbf{M}$ are sorted in increasing order. Although eqs.~(\ref{eq:boundaries1}) and~(\ref{eq:boundaries2}) do not apply to the normalised eigenvalues of $\mathbf{M}$ and $\covp$, these relations help to provide an understanding of the small number of relevant flow PCs. Both the lower bound in eq.~(\ref{eq:boundaries1}) and the upper bound in eq.~(\ref{eq:boundaries2}) correspond to the case that $\mathbf{T}$ and $\covp$ have common eigenvectors, such that the eigenvalues of $\mathbf{M}$ are just the products of the respective eigenvalues of these matrices. The lower bound is obtained if the eigenvalues occur in opposite order - that is, the eigenvector with the largest eigenvalue for $\covp$ is an eigenvector of $\mathbf{T}$ associated with its smallest eigenvalue etc. In contrast, for the upper bound the common eigenvectors are associated with eigenvalues sorted in the same order.

\begin{figure}
	\includegraphics[width=\linewidth]{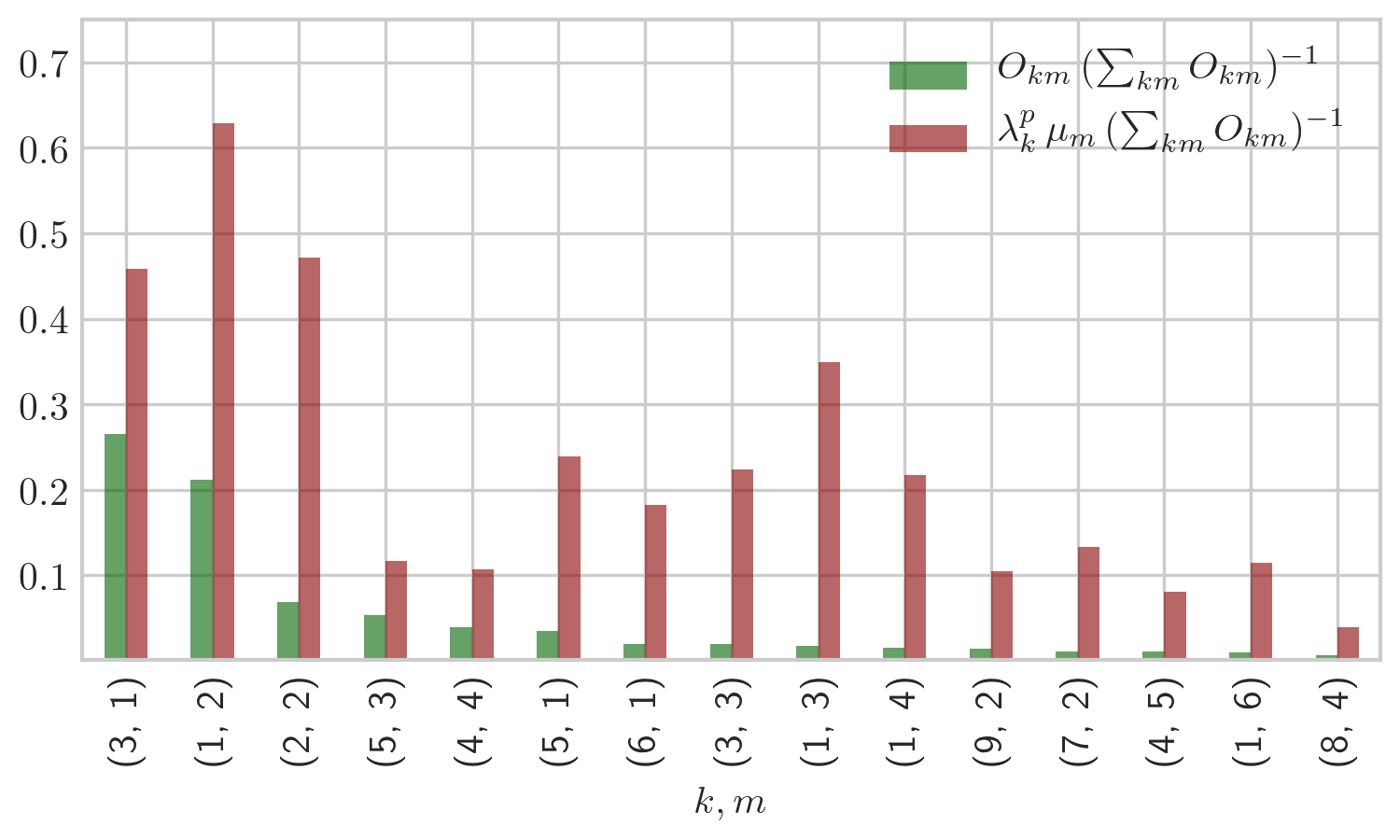}
	\caption{The green bars show the 15 largest entries of the overlap matrix $O_{km}=\lambda_{k}^p\,\mu_{m}\,(\PAp_{k}\cdot\mathbf{s}_{m})^2$, normalised by the total sum $\sum_{km}O_{km}=\text{tr}(\covf)$. The red bars show the normalised pure products $\lambda_{k}^p\,\mu_{m}(\sum_{km}O_{km})^{-1}$ corresponding to $\PAp_{k}=\mathbf{s}_{m}$.}
	\label{fig:overlapmatrix}
\end{figure}

\begin{figure}
	\includegraphics[width=\linewidth]{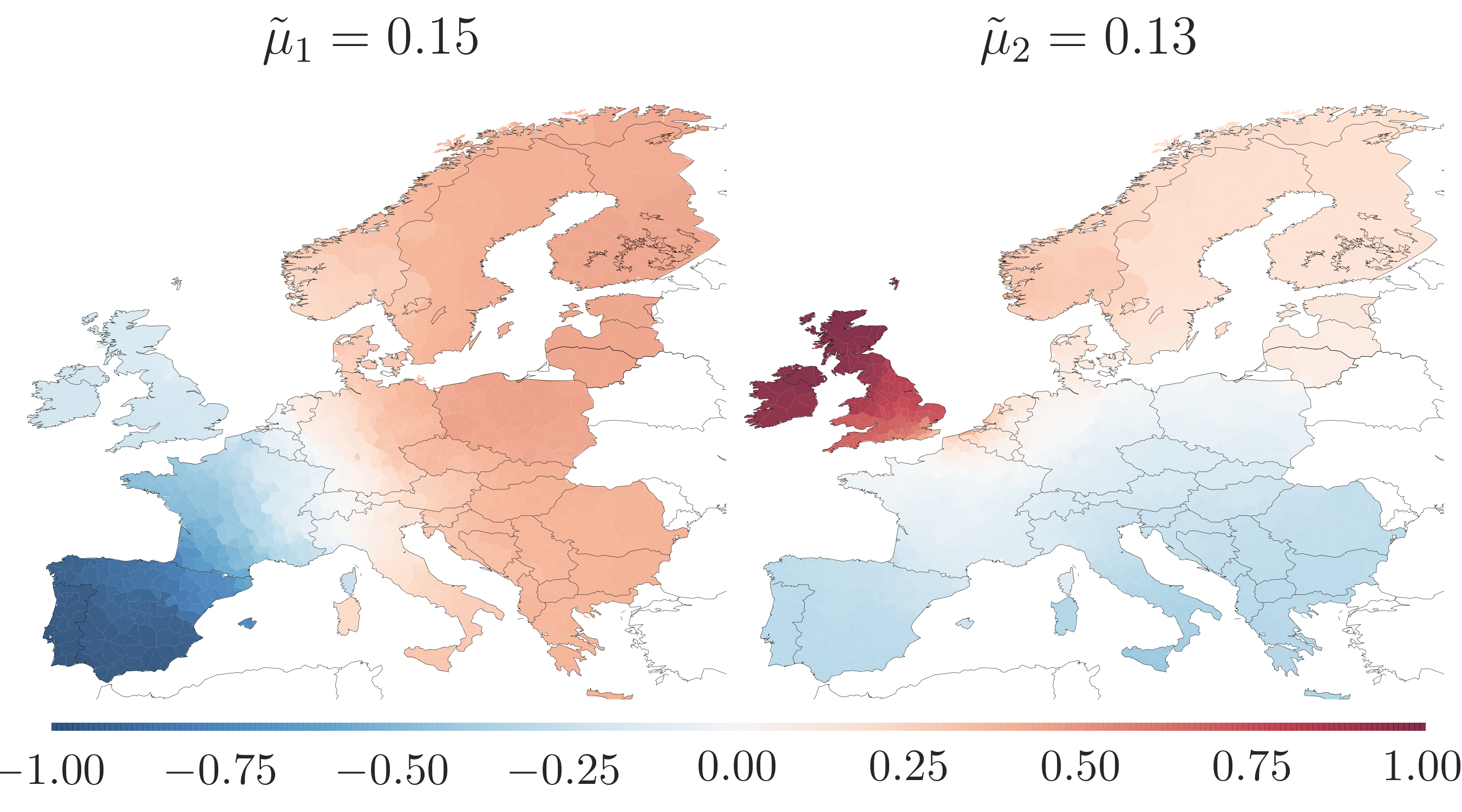}
	\caption{The eigenvectors $\mathbf{s}_{m}$ associated with the two first eigenvalues $\mu_{m}$ of the topological matrix $\mathbf{T}=\mathbf{H}^{T}\mathbf{H}$.}
	\label{fig:topology_components_pre6-1024-harm-alpha_inhom-064_v}
\end{figure}

These results relate to the finding of a small number of relevant principal flow components as follows. By definition a small number of principal flow components corresponds to the situation that the sum over the $K$ largest eigenvalues $\lambda_{k}^{f}$ at first strongly grows with $K$ and then quickly saturates. Such a behaviour is shown by the upper bound in eq.~(\ref{eq:boundaries2}) if both the distribution of the eigenvalues of $\covp$ and of $\mathbf{T}$ are heterogeneous, such that the distribution of the product of their ordered sequences displays an even higher degree of heterogeneity. This condition of a heterogeneous distribution of the eigenvalues of $\covp$ is expressed by the comparatively small number of principal injection components, whereas for the eigenvalues of $\mathbf{T}$ it follows from the network topology which can to first order be approximated as a two-dimensional lattice~\cite{schaefer2017}.
 %
 \begin{figure}
	\includegraphics[width=\linewidth]{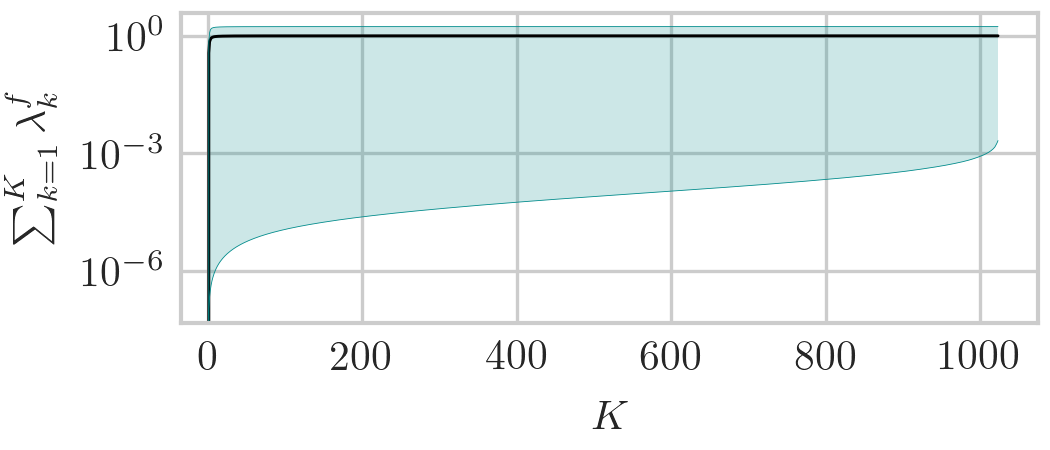}
	\caption{The black line shows the partial sum $\sum_{k=1}^{K}\lambda_{k}^{f}$ of the flow eigenvalues $\lambda_{k}^{f}=\mathbf{\Lambda}_{k}(\Sigma^{f})$ dependent on the upper limit $K=1,\ldots,N$. The shaded area visualises the upper and lower bound of this sum as given by eqs.~(\ref{eq:boundaries1}) and~(\ref{eq:boundaries2}). All terms in the figure are expressed relative to $\sum_{k=1}^{N}\lambda_{k}^{f}=\text{tr}(\mathbf{\Sigma}^{f})$.}
	\label{fig:majorization_pre6-1024-harm-alpha_inhom-064}
\end{figure}
%
 Figure~\ref{fig:majorization_pre6-1024-harm-alpha_inhom-064} shows for the detailed network representation with $N=1024$ nodes the partial sums over the eigenvalues $\sum_{k=1}^{K}\lambda_{k}^{f}$ depending on $K$, and the upper and lower bound derived from the majorisation in eqs.~(\ref{eq:boundaries1}) and~(\ref{eq:boundaries2}). As expected, the upper limit increases sharply with a high slope and already saturates at low numbers of $K$. The sum $\sum_{k=1}^{K}\lambda_{k}^{f}$ follows this behaviour close to the upper limit, which is in agreement with the small number of principal flow components displayed in fig.~\ref{fig:K_size_dependence}. It should be emphasised that both the upper and lower limit in eqs.~(\ref{eq:boundaries1}) and~(\ref{eq:boundaries2}) are expressed in terms of the eigenvalues of the matrices $\covp$ and $\mathbf{T}$. The behaviour of the sum $\sum_{k=1}^{K}\lambda_{k}^{f}$ in these bounds is thus entirely determined by the relations between the corresponding eigenvectors, which we quantify using the following overlap matrix:
 \begin{align}
	O_{km} = \lambda_{k}^p\,\mu_{m}\,(\PAp_{k}\cdot\mathbf{s}_{m})^2~.
\label{eq:overlap_matrix}
\end{align}
Here $\mathbf{s}_{m}$ denotes the (normalised) eigenvector of $\mathbf{T}$ associated with the eigenvalue $\mu_{m}$. It holds
 \begin{align}
 \sum_{k,m}O_{km} = \text{tr}(\covp \mathbf{T}) = \text{tr}(\mathbf{M}) = \text{tr}(\covf)~,
\label{eq:sum_Okm}
 \end{align}
 which already suggests that a heterogeneous distribution of $O_{km}$ can be associated with a heterogeneous distribution of the eigenvalues $\lambda_{k}^{f}$ of $\covf$. 
Figure~\ref{fig:overlapmatrix} shows the $15$ largest entries of the overlap matrix $O_{km}$ normalised by $\sum_{km}O_{km}=\text{tr}(\covf)$ in comparison to the products of the respective eigenvalues $\lambda_{k}^{p}\,\mu_{m}$. One observes a sharp decrease of these values mirroring the heterogeneous distribution of $\lambda_{k}^{f}$. The large values of $O_{km}$ correspond to an aligment of eigenvectors corresponding to large eigenvalues of $\covp$ and $\mathbf{T}$. This holds in particular for the two largest values $O_{3,1}$ and $O_{1,2}$, which together already account for $42\%$ of the total sum equal to $\text{tr}(\covf)$. Although for instance the product of the eigenvalues $\lambda_{1}^{p}\mu_{3}$ indicate a potential large contribution to the sum in eq.~(\ref{eq:sum_Okm}), the non-alignment of the corresponding eigenvectors is associated with a small value of $(\PAp_{1}\cdot\mathbf{s}_{3})^2$. Figure~\ref{fig:topology_components_pre6-1024-harm-alpha_inhom-064_v} shows the  eigenvectors associated with the two first eigenvectors of $\mathbf{T}=\mathbf{H}^{T}\mathbf{H}$. The similarity of these patterns with the third and first principal injection component shown in fig.~\ref{fig:injection_components_pre6-1024-harm-alpha_inhom-064}, respectively, explains the high values of the corresponding entries of the overlap matrix. Whereas for the injection patterns $\covp$ the eigenvectors are connected to specific properties of the system (weather and load patterns as well as the balancing mechanism), for the topological matrix $\mathbf{T}$ the eigenvalues can be understood from the network structure as follows. For unit line susceptances, $\mathbf{\Omega}$ is the identity matrix, the nodal susceptance matrix $\mathbf{B}$ is equal to the network Laplacian $\mathbf{L}$,  and the topological matrix $\mathbf{T}$ can be written as the Pseudo-Inverse of the Laplacian, $\mathbf{T}=\mathbf{H}^{T}\mathbf{H}=\mathbf{L}^{+}$~\cite{gutman1996,schaefer2017}. In this approximation the eigenvectors of $\mathbf{T}$ correspond to the eigenvectors of the network Laplacian. If we consider the network structure of the power grid to first order as a two-dimensional lattice, these topological eigenvectors~$\mathbf{s}_{m}$ can be approximated as superpositions of sine-waves along the north-south and the east-west direction with increasing frequency~\cite{vanmieghem2010}. In particular, this structure is visible in fig.~\ref{fig:topology_components_pre6-1024-harm-alpha_inhom-064_v} as the two first non-homogeneous modes of the network structure. 
 %
%
%
%
%
%
\section{Conclusion}
%
%
In this work the principal power injection and flow patterns for a model of a highly renewable European electricity network are presented. By applying PCA and investigating the spatio-temporal behavior of the injection components, it is shown that the most important recurring patterns are strongly weather-driven. The first principal injection patterns are roughly shaped as gradients along geographical axes (north-south, east-west), as the energy supply is subject to strong wind power fluctuations in North Europe and solar power variance in the south. At high network resolution, wind power generation areas along the North Sea coast as well as load centers with densely populated areas are highlighted. The number of relevant injection components $ \mathcal{K} $ at first scales roughly with the logarithm of the spatial scale of the network, but then saturates at $N=512$ nodes. In contrast, turning from the nodal power injections to the transmission line flows, the power flow data in the whole range from low ($ N=37 $) to high network resolutions ($ N=1024 $) can accurately be reproduced by around eight components. The corresponding long-distance flow patterns, which cross the network with homogeneous orientation, result from an alignment of the power injection patterns and the network topology. By defining an overlap matrix, we observe that the injection covariance matrix and the newly introduced topological matrix match in a very small subspace which includes the most important eigenvectors. This leads to a boost of fluctuations on a small set of principal flow patterns and lifts their eigenvalues.

These observations suggest that transmission studies might be greatly simplified. Instead of analysing large ensembles of possible flow situations, for most purposes network planners can focus on the few patterns that determine the majority of flows in the majority of situations. Furthermore, these patterns are robust against changes in the network resolution.

The results of this work however lead to further research questions. The presented model assumes unconstrained transmission capacities as well as a simplified balancing scheme. It would be interesting to investigate the extent to which the findings hold for an optimised dispatch with transmission constraints. For this case one could expect an increase in the number of relevant flow components as the long-distance patterns might be suppressed.
%
%
\acknowledgments
%
%
The authors thank Johannes Kruse for fruitful discussions. M. S. is partially funded by the Carlsberg Foundation Distinguished Postdoctoral Fellowship. M. G. is partially funded by the RE-INVEST project (Renewable Energy Investment Strategies -- A two-dimensional interconnectivity approach), which is supported by Innovation Fund Denmark (6154-00022B). T.B. and J.H. acknowledge funding from the Helmholtz Association under grant no.
VH-NG-1352 and from the German Federal Ministry of Education and
Research (BMBF) under grant no. 03SF0472C The responsibility for the contents lies solely with the authors.
%
%
\bibliographystyle{unsrt}
\bibliography{references}
%
%
\end{document}